\documentclass[letterpaper, 10 pt, conference]{ieeeconf}  % Comment this line out if you need a4paper

\usepackage{blindtext}
\usepackage{eso-pic}
\usepackage{amsmath}
\usepackage{booktabs}
\usepackage{cite}
\usepackage{amsmath,amssymb,amsfonts}
\usepackage{algorithmic}
\usepackage{graphicx}
\usepackage{textcomp}
\usepackage{xcolor}
\usepackage{hyperref}
\usepackage{multirow}
\usepackage{booktabs}

\IEEEoverridecommandlockouts                              % This command is only needed if 
\title{Ensemble Models for Predicting Treatment Response in Pediatric Low-Grade Glioma Managed with Chemotherapy}

\author{Max Bengtsson$^{1}$, Elif Keles$^{1}$, Angela J. Waanders$^{1,2}$ and Ulas Bagci$^{1}$%
\thanks{*This work is supported by Malnati Brain Tumor Initiative, Northwestern University.}% <-this % stops a space
\thanks{$^{1}$Northwestern University, Chicago IL, USA {\tt\small maxbengtsson2025@u.northwestern.edu}}%
\thanks{$^{2}$Ann \& Robert H Lurie Children’s Hospital of Chicago, Chicago, IL, USA}%
}

\begin{document}

\maketitle
\thispagestyle{empty}
\pagestyle{empty}

%%%%%%%%%%%%%%%%%%%%%%%%%%%%%%%%%%%%%%%%%%%%%%%%%%%%%%%%%%%%%%%%%%%%%%%%%%%%%%%%
\begin{abstract}

In this paper, we introduce a novel pipeline for predicting chemotherapy response in pediatric brain tumors that are not amenable to complete surgical resection, using pretreatment magnetic resonance imaging combined with clinical information. Our method integrates a state-of-the-art pediatric brain tumor segmentation framework with radiomic feature extraction and clinical data through an ensemble of a \textbf{Swin UNETR} encoder and \textbf{XGBoost} classifier. The segmentation model delineates four tumor subregions enhancing tumor (ET), non-enhancing tumor (NET), cystic component (CC) and edema (ED) which are used to extract imaging biomarkers and generate predictive features. The \textbf{Swin UNETR} network classifies the response to treatment directly from these segmented MRI scans, while \textbf{XGBoost} predicts response using radiomics and clinical variables including legal sex, ethnicity, race, age at event (in days), molecular subtype, tumor locations, initial surgery status, metastatic status, metastasis location, chemotherapy type, protocol name and chemotherapy agents. The ensemble output provides a  non-invasive estimate of chemotherapy response in this historically challenging population characterized by lower progression-free survival. Among compared approaches, our \textbf{Swin-Ensemble} achieved the best performance (precision for non effective cases=0.68, recall for non effective cases=0.85, precision for chemotherapy effective cases=0.64 and overall accuracy=0.69), outperforming \textbf{Mamba-FeatureFuse}, \textbf{Swin UNETR} encoder, and \textbf{Swin-FeatureFuse} models. Our findings suggest that this ensemble framework represents a promising step toward personalized therapy response prediction for pediatric low-grade glioma patients in need of chemotherapy treatment who are not suitable for complete surgical resection, a population with significantly lower progression free survival and for whom chemotherapy remains the primary treatment option. Our model is available here: \href{https://github.com/NUBagciLab/pediatric-chemotherapy-response-prediction}{Github}.

\end{abstract}


%%%%%%%%%%%%%%%%%%%%%%%%%%%%%%%%%%%%%%%%%%%%%%%%%%%%%%%%%%%%%%%%%%%%%%%%%%%%%%%%
\section{INTRODUCTION}

Pediatric Central Nervous System (CNS) tumors present significant clinical challenges, as they are the most common solid tumors affecting children and adolescents, ranking second to leukemia in overall frequency\cite{girardi2023global,fangusaro2024pediatric}.  Critically, pediatric brain tumors are the leading cause of cancer-related mortality in this population, underscoring the urgent need for improved diagnostic and prognostic strategies\cite{girardi2023global}.
Surgical resection is the mainstay of front-line therapy for symptomatic pLGG; however, more than 50\% of pLGGs occur in locations that are either not amenable to surgery or only amenable to a biopsy or limited resection \cite{zapotocky2024low}. Tumors  typically cannot be fully removed and therefore rely on chemotherapy as first-line treatment, yet five-year progression-free survival (PFS) in these patients remains only about 45 to 55 percent compared to the surgically removed cases of 85 to 90 percent \cite{zapotocky2024low}. Outcomes are even poorer in tumors with high-risk molecular features such as BRAF V600E mutations or CDKN2A deletions, where long-term PFS has been reported as low as 27 percent compared to 60 percent without the same variant \cite{cipri2023unlocking}. These children frequently experience repeated relapses and substantial visual, endocrine and neurocognitive morbidity, highlighting the aggressive behavior and chronic nature of pLGG in this high-risk population \cite{sait2023treatment}.

Accurate prediction of tumor characteristics and clinical outcomes plays a pivotal role in optimizing patient management. Current diagnostic pathways rely heavily on Magnetic Resonance Imaging \cite{girardi2023global} and clinical indicators\cite{grudzien2023predicting,vihermakimachine,zhang2022mri,bareja2024nnu}. To meet these clinical needs, recent advances in quantitative imaging, radiomics, and Artificial Intelligence (AI) particularly Machine Learning and Deep Learning (DL) have opened new avenues for non-invasive prediction of diagnostic, prognostic, and molecular features \cite{familiar2024towards,elif2025comment,pacchiano2024radiomics,grudzien2023predicting}. These approaches analyze high-dimensional MRI data to extract imaging biomarkers that correlate with tumor biology and clinical outcomes \cite{familiar2024towards,elif2025comment,bareja2024nnu}.

Recent studies have explored diverse prediction tasks across diagnostic, prognostic, and molecular domains \cite{vihermakimachine}. Prognostic frameworks combining imaging and clinical data have stratified patients by overall survival (OS), PFS, and event-free survival (EFS) and predicted molecular subtype \cite{familiar2024towards,vihermakimachine,mahootiha2025multimodal}. However, despite these advances, no studies have yet focused on predicting therapy response at the time of diagnosis, especially the shorter overall survival group of chemotherapy treated pLGG patients. Early identification of treatment responsiveness is essential for achieving individualized therapy, enabling clinicians to tailor treatment and aim for better overall survival with current treatment options.
To achieve these unmet clinical needs, we aimed to predict chemotherapy response in low grade pediatric glioma patients (pLGG) age ranges from 129 to 7019 days.
Our contributions are summarized as follows:
\begin{enumerate}
    \item We developed a pipeline of Swin UNETR encoder and XGBoost ensemble to predict therapy response using only pretreatment imaging and a small amount of clinical information. 
    \item We utilize the SOTA pediatric brain tumor segmentation model, built on radiological reasoning, and its prediction masks to enable our pipeline.
    \item Our model is the first to predict therapy response in this clinically challenging group of pLGG patients treated with chemotherapy.
\end{enumerate}

\section{RELATED WORKS}

Recent research has increasingly demonstrated that combining quantitative MRI features with clinical variables can substantially improve prognostic stratification in pLGGs\cite{wagner2021radiomics,liu2023radiomic,fathi2025multiparametric}. In a multi-institutional exploratory radiomics study, high-dimensional texture and shape features extracted from pretreatment MRI were shown to non-invasively classify molecular risk in pLGGs\cite{wagner2021radiomics,liu2023radiomic}.  Another study integrated imaging, transcriptomics (immune-clusters) and clinical data, achieving a C-index of ~0.71–0.77 for PFS prediction\cite{fathi2025multiparametric}. Two independent studies in large multi-institutional cohorts showed that deep learning models integrating MRI-derived imaging features with clinical variables substantially improved prediction of EFS in pediatric low-grade glioma, outperforming clinical-only and imaging-only approaches and enabling more accurate risk stratification\cite{mahootiha2025multimodal,zapaishchykova2023lgg}. These studies underscore that while traditional clinical features remain foundational in prognostication, the incorporation of advanced imaging biomarkers and machine-learning techniques offers meaningful incremental value. Most existing studies focus on survival or progression-related outcomes rather than on predicting chemotherapy response in pLGG patients who are not suitable candidates for complete surgical resection and therefore require chemotherapy as their primary treatment modality. This clinical challenge highlights a significant need for further research in individualized therapy planning. In our study, we introduce an algorithm that predicts treatment response, providing a near-term, outcome-specific measure that more accurately captures the biological and clinical determinants of chemotherapeutic success. This strategy enables clinicians to identify likely responders at the time of diagnosis and consider alternative regimens for predicted non-responders, ultimately supporting more individualized care pathways.

\section{METHODOLOGY}

\subsection{Dataset Creation}

\begin{figure*}[htbp]
    \centering
    \includegraphics[width=0.9\textwidth]{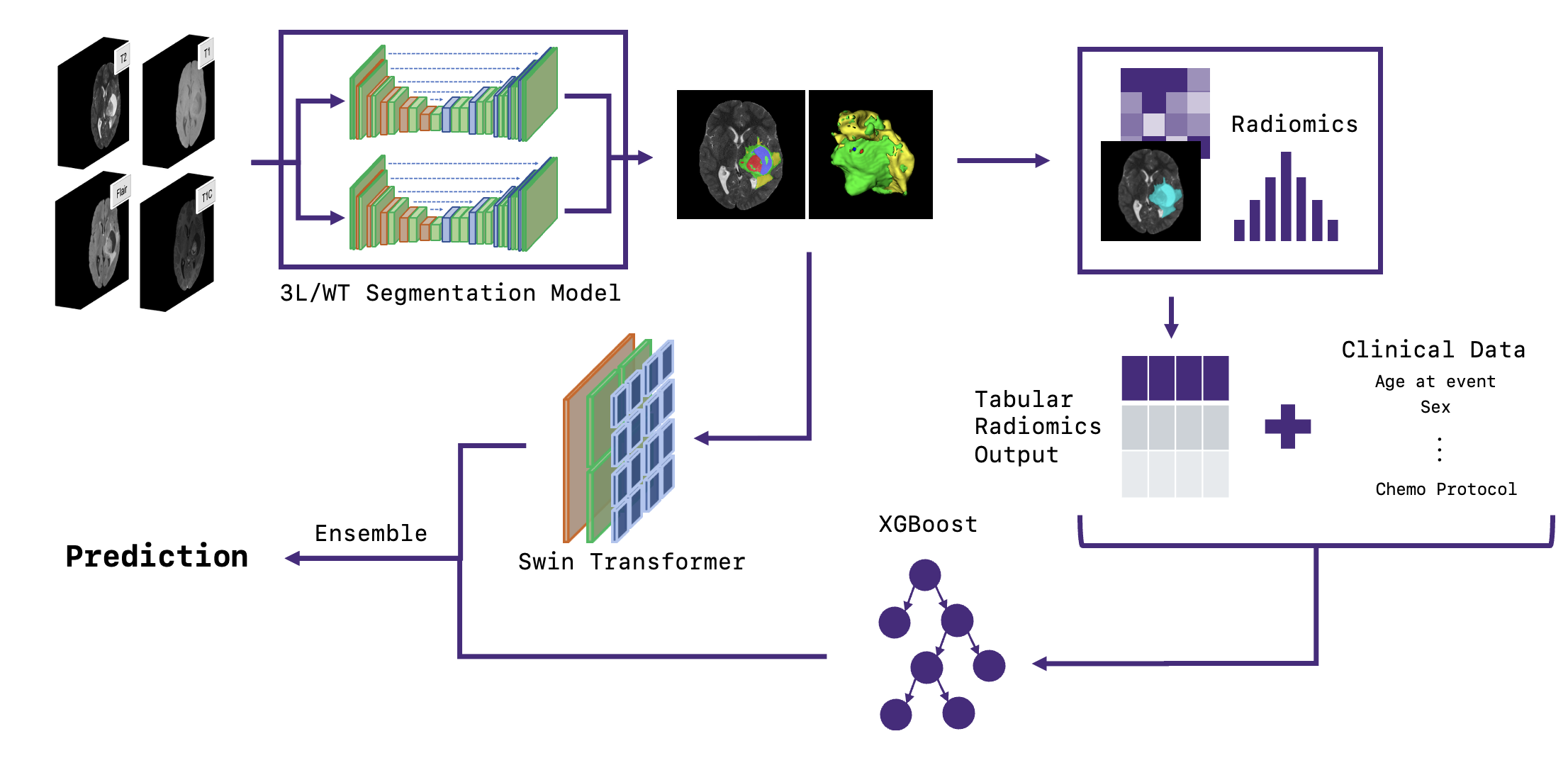}
    \caption{Classification Pipeline Architecture Including Segmentation, Radiomic Feature Extraction, Clinical Data Combination and Classification.}
    \label{fig:pipeline}
\end{figure*}

Model training and testing utilizes the Children’s Brain Tumor Network’s (CBTN) dataset, with our version consisting of 622 Patients \cite{lilly2023children}. Of these 622 patients we selected only patients who had received chemotherapy. A pediatrician and radiologist examined each remaining patient’s initial acquisitions and selected cases where the T1, T1 Contrast Enhanced (T1CE), T2 and T2 FLAIR (FLAIR) sequences were present. Every patient's images are preprocessed using the 4 selected sequences through the CaPTK preprocessing pipeline \cite{pati2020cancer,davatzikos2018cancer}. This includes reorientation to LPS/RAI, registration of T1CE to SRI-24 atlas, registration of T1, T2 and FLAIR to T1CE and N4 Bias Correction. Skull stripping the preprocessed images is done using an nnU-Net based skull stripping model \cite{vossough2024training,isensee2021nnu}.

The aforementioned selection process resulted in a dataset of 105 patients (aged 129 to 7019 days) which were then labeled and used to conduct therapy response prediction. Labeling was conducted using clinical information by comparing the dates of patient's chemotherapy timeline and its relation to progression or recurrence events. It was therefore necessary to exclude certain patients where their start or end of chemotherapy was not noted. For the remaining patients, we compared the age at the first Event Free Survival (EFS) event, start of chemotherapy and end of chemotherapy. Here EFS event is defined as the earliest event after the initial CNS tumor diagnosis which indicates disease progression, recurrence or a second malignancy. Our labeling is expressed in \eqref{eq:outcome_definition} where \(T_{chemo_{start}}\) represents age in days of starting chemotherapy, \(T_{chemo_{end}}\) represents age in days of ending chemotherapy, \(T_{EFS}\) is the age in days at EFS event and where \(0\) is labeled as not effective.

\begin{equation}
\label{eq:outcome_definition}
\text{Outcome} = 
\begin{cases} 
  0 & \text{if deceased due to illness} \\
  1 & \text{if \(T_{EFS} = NA\)} \\
  1 & \text{if \(T_{EFS} < T_{chemo_{start}}\)} \\
  0 & \text{if \(T_{chemo_{start}} < T_{EFS} < T_{chemo_{end}}\)} \\
  0 & \text{if \(T_{chemo_{end}} < T_{EFS}\)}
\end{cases}
\end{equation}

Patient's who are deceased due to illness at last known clinical status are labeled as not effective (7 patients), and if a patient does not exhibit an EFS event they are labeled as effective (41 patients). For the remaining patients we compare the timing of the EFS event relative to their start and stop of chemotherapy. If the first EFS event occurs before beginning chemotherapy and there were no following EFS events then we classify these patients as effective (1 patient). If the event occurs either during chemotherapy or after chemotherapy has been completed then this is labeled as not effective (56 patients). This labeling results in a distribution of 42 patients being labeled effective and 63 being labeled as not effective.

\subsection{Classification Pipeline Architecture}

Our model predicts outcome with simply inputs of pretreatment imaging and basic clinical information from the patient. The implementation as shown in Figure \ref{fig:pipeline} uses three different stages: segmentation, radiomic feature extraction, and classification. Firstly, segmentation of the tumor is conducted with an nnU-Net based state-of-the-art (SOTA) pediatric brain tumor segmentation model which generates different subregions of the tumor mask separately \cite{bengtsson2025new}. This model tested on a CBTN test set of 30 patients scored a 68.1\% dice outperforming the previous SOTA scoring 65.9\%. This pretrained algorithm segments the entire tumor using two distinct nnU-Net models. One of the models segment the enhanced tumor (ET), cystic component (CC) and edema (ED) while the other exclusively segments the whole tumor (WT). The outputs are then combined into a single 4 label output where any region not classified as ET, CC or ED but classified as WT is reclassified to non-enhanced tumor (NET). This processes is described in \eqref{eqn:yNet} and \eqref{eqn:finalSeg} where \(y_{NET}\) is the label of NET and \(\mathcal{S}_{final}\) is the final 4 label segmentation mask.

\begin{equation}
\label{eqn:yNet}
    y_{NET} = y_{WT} \setminus (y_{ET} \cup y_{CC} \cup y_{ED}).
\end{equation}
\begin{equation}
\label{eqn:finalSeg}
    \mathcal{S}_{final} = \{ y_{ET}, y_{CC}, y_{ED}, y_{NET} \}.
\end{equation}

The segmentation output of ET, NET, CC and ED is then recombined into a new combined WT mask to be used with radiomic feature extraction and seen in figure \ref{fig:pipeline} \cite{van2017computational}. In conjunction with the WT mask, we use the T2 MRI image as input for radiomic feature extraction and process all 105 patients. The output is tabular with 102 total features per patient, including shape, first order features, gray level co-occurrence and more. The tabular output is then concatenated with the patient’s clinical data before being passed on to the classifier with an example shown in table \ref{tab:tabularDataExample}. The following clinical data is used during this process: legal sex, ethnicity, race, age at event (in days), molecular subtype, tumor locations, initial surgery status, metastatic status, metastasis location, chemotherapy type, protocol name and chemotherapy agents.

\begin{table}[h!]
\centering
\caption{Example of Combined Tabular Data Used by XGBoost.}
\label{tab:tabularDataExample}

% --- Table Definition ---
% We define 6 columns (4 'c' and 2 'c')
\begin{tabular}{cccc cc}
\toprule

% --- Group Headers (Row 1) ---
% \multicolumn{<num_cols_to_span>}{<alignment>}{<text>}
% This row has TWO items (two multicolumns)
\multicolumn{4}{c}{\textbf{Radiomic Features (102 cols)}} & \multicolumn{2}{c}{\textbf{Clinical Data (12 cols)}} \\

% --- Partial Rules (Row 2) ---
% \cmidrule(lr){<start_col>-<end_col>}
% (r) trims the right, (l) trims the left. This keeps them from merging.
\cmidrule(r){1-4} \cmidrule(l){5-6}

% --- Individual Column Headers (Row 3) ---
% This row has 6 items, separated by 5 ampersands (&)
Elongation & Flatness & \dots & Zone Var & Age (Days) & \dots \\
\midrule

% --- Data Row 1 (Row 4) ---
% This row also has 6 items, separated by 5 ampersands (&)
0.863 & 0.692 & \dots & 117.12 & 2973 & \dots \\

% --- Data Row 2 (Row 5) ---
0.723 & 0.587 & \dots & 209.31 & 2398 & \dots \\
\bottomrule
\end{tabular}
\end{table}

\begin{table*}[htbp]
    \centering
    \caption{Classification results of models where P is precision, R is recall, 0 is not effective and 1 is effective.}
    \label{tab:scores}
    \begin{scriptsize}  % Start of font size 8
    \begin{tabular}{|c|c|c|c|c|c|c|c|c|c|c|}
        \hline
        Model & P - 0 & R - 0 & F1 - 0 & P - 1 & R - 1 & F1 - 1 & Accuracy & P - Avg. & R - Avg. & F1 - Avg. \\
        \hline
        XGBoost&0.66±0.21&0.68±0.12&0.66±0.16&0.49±0.17&\textbf{0.49±0.20}&0.47±0.17&60.95±12.92&0.57±0.11&0.58±0.12&0.56±0.12 \\
        Conv-Deep Learning&0.62±0.14&0.76±0.22&0.64±0.04&0.23±0.28&0.27±0.23&0.23±0.23&52.38±6.73&0.42±0.15&0.51±0.07&0.43±0.11 \\
        Conv-Ensemble&0.68±0.18&0.79±0.09&0.71±0.12&0.55±0.19&0.44±0.18&0.46±0.16&64.76±11.89&0.61±0.10&0.61±0.08&0.59±0.10 \\
        Conv-FeatureFuse&0.67±0.16&0.83±0.15&0.71±0.06&\textbf{0.66±0.28}&0.42±0.17&0.42±0.11&61.91±6.73&\textbf{0.67±0.06}&0.62±0.03&0.57±0.07 \\
        Mamba-Deep Learning&0.66±0.20&0.76±0.23&0.65±0.11&0.51±0.37&0.38±0.34&0.33±0.20&55.24±11.11&0.58±0.17&0.57±0.11&0.49±0.12 \\
        Mamba-Ensemble&0.63±0.17&0.81±0.17&0.68±0.11&0.45±0.30&0.32±0.21&0.31±0.17&59.05±8.83&0.54±0.11&0.56±0.05&0.50±0.06 \\
        Mamba-FeatureFuse&0.62±0.15&0.76±0.17&0.65±0.04&0.42±0.33&0.29±0.23&0.27±0.19&53.33±4.67&0.52±0.13&0.52±0.10&0.46±0.09 \\
        Swin-Deep Learning&0.64±0.15&0.77±0.17&0.67±0.08&0.56±0.30&0.35±0.19&0.35±0.15&58.09±6.32&0.60±0.08&0.56±0.03&0.51±0.07 \\
        \textbf{Swin-Ensemble}&\textbf{0.68±0.18}&\textbf{0.85±0.07}&\textbf{0.75±0.13}&0.64±0.21&0.44±0.14&\textbf{0.51±0.16}&\textbf{68.57±12.99}&0.66±0.12&\textbf{0.64±0.10}&\textbf{0.63±0.12} \\
        Swin-FeatureFuse&0.58±0.10&0.72±0.23&0.61±0.10&0.49±0.36&0.21±0.12&0.27±0.15&49.52±11.51&0.54±0.14&0.47±0.16&0.44±0.12 \\
        \hline
    \end{tabular}
    \end{scriptsize}  % End of font size 8
\end{table*}

The classifier layer consists of an ensemble between an XGBoost model trained and tested on the tabular radiomics and clinical dataset shown in table \ref{tab:tabularDataExample} and a deep learning based classifier acting on the 4 label segmentation, \(\mathcal{S}_{final}\), and 4 sequence imaging \cite{Chen_2016}. Multiple different deep learning classifiers built on different backbones including a convolutional ResNet style encoder, mamba encoder, and a Swin UNETR encoder where tested \cite{lecun2002gradient,he2016deep,xing2024segmamba,hatamizadeh2021swin,liu2021swin}. Training and testing images and masks were cropped with a 5 pixel margin around the tumor masks and then resized to 64x64x64. During training random rotation and flipping was added to both the masks and images in addition to gaussian noise being applied to the images.

When benchmarking we used two architecture styles. The first one combined deep learning with XGBoost through ensembling final predictions. The second used feature extraction of the deep learning models concatenated with radiomics and clinical tabular data  to train and test with XGBoost which are labeled FeatureFuse in table \ref{tab:scores}. Ultimately, the best performing model uses ensembling between XGBoost and the Swin UNETR encoder. This is conducted through taking the mean probabilities from the final output layer for the deep learning implementation and XGBoost as seen in \eqref{eqn:ensemble}.

\begin{equation}
\label{eqn:ensemble}
    g(x_{i,complete}) = \frac{f(x_{i,imaging}) + h(x_{i,tabular})}{2},
\end{equation}
here \(g(x)\) represents the model ensemble prediction using \(x_{complete}\), the imaging, segmentation and tabular data. \(f(x)\) and \(h(x)\) represent the deep learning prediction using the imaging and segmentation data \(x_{imaging}\) and XGBoost using the tabular data \(x_{tabular}\) respectively. 

When benchmarking FeatureFuse models we use a hook to extract weights from the penultimate layer before the classification head. The weights are concatenated with the radiomic features and clinical data which are trained and tested with XGBoost as a new dataset. Although sometimes improving performance generally these did not perform as well as XGBoost or deep learning models themselves and were outperformed by ensembling models.

\begin{table}[h!]
\centering
\caption{Comparision of random forest, SVM and XGBoost}
\label{tab:compareML}

% --- Table Definition ---
% We define 6 columns (4 'c' and 2 'c')
\begin{tabular}{|c|c|c|c|}
\hline
Model & Train Acc. & Validation Acc. & Test Acc. \\
\hline
Random Forest & 97.81±1.10 & 69.09±12.33 & 55.24±7.13 \\
SVM & 47.67±4.01 & 72.73±8.13 & 58.10±7.73 \\ 
XGBoost & 100.0±0.0 & 72.73±5.75 & 60.95±12.92 \\
\hline

\end{tabular}
\end{table}

\begin{table}[h!]
\centering
\caption{Comparison of XGBoost training using only clinical data, only radiomic data or combined data}
\label{tab:compareClinical}

% --- Table Definition ---
% We define 6 columns (4 'c' and 2 'c')
\begin{tabular}{|c|c|c|c|}
\hline
Data & Train Acc. & Validation Acc. & Test Acc. \\
\hline
Clinical & 70.41±1.86 & 63.64±12.86 & 61.90±13.13 \\
Radiomics & 78.63±4.71 & 65.45±10.60 & 56.19±8.19 \\ 
Combined& 100.0±0.0 & 72.73±5.75 & 60.95±12.92 \\
\hline

\end{tabular}
\end{table}

Selecting XGBoost as our model classifier is supported by table \ref{tab:compareML} where we benchmark XGBoost against random forest and SVM \cite{breiman2001random,cortes1995support}. XGBoost is investigated further through analyzing the performance of models trained independently on just clinical or radiomic information and comparing this to a full model trained on the combined data as seen in table \ref{tab:compareClinical}. On these same models we further evaluate the average score of the 10 best parameter selections which can be seen in table \ref{tab:top10avg}. Feature analysis of our full XGBoost model using SHAP is then conducted and we display the absolute value of respective SHAP values across the dataset on a 5 fold basis. The results are found in table \ref{tab:SHAP}.

\begin{table}[h!]
\centering
\caption{Comparison of clinical, radiomics and combined data by taking the average testing score of the top 1\% best scoring parameters on the validation set}
\label{tab:top10avg}

% --- Table Definition ---
% We define 6 columns (4 'c' and 2 'c')
\begin{tabular}{|c|c|}
\hline
Data & Avg. Test Acc. (Top 10) \\
\hline
Clinical & 61.90±0.85 \\
Radiomics & 52.86±1.87 \\ 
\textbf{Combined}& \textbf{63.24±1.55} \\
\hline

\end{tabular}
\end{table}

\begin{table}[h!]
\centering
\caption{SHAP feature importance}
\label{tab:SHAP}

% --- Table Definition ---
% We define 6 columns (4 'c' and 2 'c')
\begin{tabular}{|c|c|}
\hline
Feature	& SHAP \\
\hline
age\_at\_event\_days & 0.99±0.17 \\
tumor\_locations	& 0.33±0.25 \\
glszm\_SizeZoneNonUniformityNormalized & 0.31±0.14 \\
firstorder\_Minimum & 0.30±0.13 \\
firstorder\_Kurtosis & 0.25±0.14 \\
firstorder\_Skewness & 0.24±0.15 \\
shape\_Flatness & 0.23±0.14 \\
shape\_Sphericity & 0.23±0.12 \\
shape\_Elongation & 0.19±0.12 \\
gldm\_DependenceEntropy	& 0.18±0.1 \\
shape\_MinorAxisLength & 0.18±0.14 \\
glcm\_Idmn & 0.16±0.11 \\
glszm\_GrayLevelNonUniformity & 0.16±0.11 \\
gldm\_LargeDependenceHighGrayLevelEmphasis & 0.16±0.10 \\
shape\_Maximum2DDiameterSlice & 0.15±0.16 \\
glszm\_ZoneEntropy & 0.15±0.11 \\
firstorder\_10Percentile & 0.14±0.07 \\
shape\_Maximum2DDiameterRow & 0.14±0.10 \\
shape\_Maximum3DDiameter & 0.14±0.10 \\
\hline

\end{tabular}
\end{table}

\subsection{Hyperparameters}

All deep learning models were trained using the ADAM optimizer with a learning rate of 1e-4. Each model was trained for 500 epochs and the model weights used for inference were selected by taking the highest score on a held out validation set intermittently tested during training. We trained with a batch size of 8 using binary cross entropy loss. Our ResNet style convolutional model was trained with depths between downsampling layers of [4,4,8,4] and feature sizes for these layers of [16, 32, 64, 128]. For the mamba model we utilize the mamba encoder from SegMamba with depths of [2, 2, 2, 2] and features of size [48, 96, 192, 384]. Lastly, for the Swin UNETR encoder we utilize depths of [2, 2, 2, 2] and [3, 6, 12, 24] number of heads.

Our XGBoost model parameters were found using a bounded random parameter search with ranges specified in table \ref{tab:xgboostParams}. Using a uniform distribution over these specified ranges we samples parameters with integer values for max depth and min child weight, floating point values rounded to 2 decmial places for subsamples, colsample by tree and learning rate. Alphas, lambdas and gammas where selected as floating point values rounded to one decimal. We sampled the space 1000 times resulting in 3 models with the the highest validation score of 72.73\%. Testing these three we choose the model with the smallest max depth of 3. All FeatureFuse models were sampled similarly using 1000 samples of the parameter space with changes to parameter ranges reflecting the following: max depth (3...8), min child weight (2...14), samples and colsample by tree (0.6...1.0), learning rate (0.01...0.3).

\begin{table}[h!]
\centering
\caption{Parameter space sampled when training XGBoost.}
\label{tab:xgboostParams}

% --- Table Definition ---
% We define 6 columns (4 'c' and 2 'c')
\begin{tabular}{cc}
\toprule
Parameter & Range \\
\midrule
Max Depths & [2,7] \\
Min Child Weights & [1,6] \\ 
Subsamples & [0.6,0.85] \\ 
Colsample by Trees & [0.6,0.85] \\ 
Learning Rates & [0.01,0.05] \\ 
Alphas & [0.0,4.0] \\ 
Lambdas & [0.0,4.0] \\ 
Gammas & [0.0,0.5] \\ 
\bottomrule

\end{tabular}
\end{table}

We additionally benchmark XGBoost against SVM and random forest classifiers fitted using a parameter search similar to XGBoost. We sampled the parameter space 1000 times and selected the best performing model on the validation set. For SVM our parameter ranges were: C (0.1...100.0), gamma (0.001...1.0), kernel (linear, rbf, poly, sigmoid), degree (2...5). Random forest parameters were drawn from the following distribution: number of estimators (100...1000), max depth (5...50), min samples per split (2...20), min samples per leaf (1...10). This same methodology carries over to the testing of clinical, radiomic and combined XGBoost models. These models were all trained using 1000 samples of the parameter space with ranges corresponding to \ref{tab:xgboostParams}.

\subsection{Training}

Training was conducted using 5 fold cross validation on the 105 patients. We separate the dataset into 70\% training, 10\% validation and 20\% testing for each of these 5 folds with every patient appearing once in the testing set. XGBoost and the individual deep learning models are then trained and tested on these folds respectively. All deep learning models were trained independently of the XGBoost models and then combined during ensembleing or feature fusion. Training was conducted on a single NVIDIA A6000 GPU with 48 GB of VRAM. 

\section{RESULTS}

The performance of different classification models can be seen in table \ref{tab:compareML} with listed scores calculated with 5 fold cross validation on training set, validation set and test set. All model's hyperparameters have been fine tuned to increase score on the validation dataset and the results show that both XGBoost and SVM outperforms random forest by 3.64\%. XGBoost and SVM do both perform well on the validation set with scores of 72.73\% however SVM does not seem to have learned the dataset since the training score is incredibly low at 47.67\% compared to XGBoost which perfectly learned the training set. Our choice of XGBoost was further supported by the test scores where XGBoost outperforms both SVM and random forest with a test set score of 60.95\%.

Final model metrics were evaluated using 5 fold cross validation where each test set consisted of 21 patients. The results can be seen in table \ref{tab:scores}. In our testing, XGBoost outperformed all the basic deep learning models by 8.57\%, 5.71\% and 2.86\% for convolutional, mamba and swin networks respectively. Across the board, we see that ensemble models performed better than FeatureFuse and individual deep learning models even when running individual parameter searches for all FeatureFuse variants. Convolutional ResNet ensemble saw a huge increase of 12.38\% accuracy over the baseline model, Mamba’s score likewise increased by 3.81\% and lastly Swin’s score also drastically increased by 10.48\%. These exhibit clear improvements in accuracy and show the importance of clinical information when classifying therapy outcomes for patients over just using imaging. FeatureFuse on the other hand sometimes saw drastic decreases in scores relative to the baseline models with decreases for mamba and Swin of 1.91\% and 8.57\% respectively. Convolutional FeatureFuse however improved on the baseline XGBoost in all metrics barring recall and F1 score for the effective class. Overall, the best performing model was Swin-Ensemble which had the best accuracy, recall and F1 score while convolutional FeatureFuse scored slightly better in precision. 

\begin{figure}[htbp]
    \centering
    \includegraphics[width=\columnwidth]{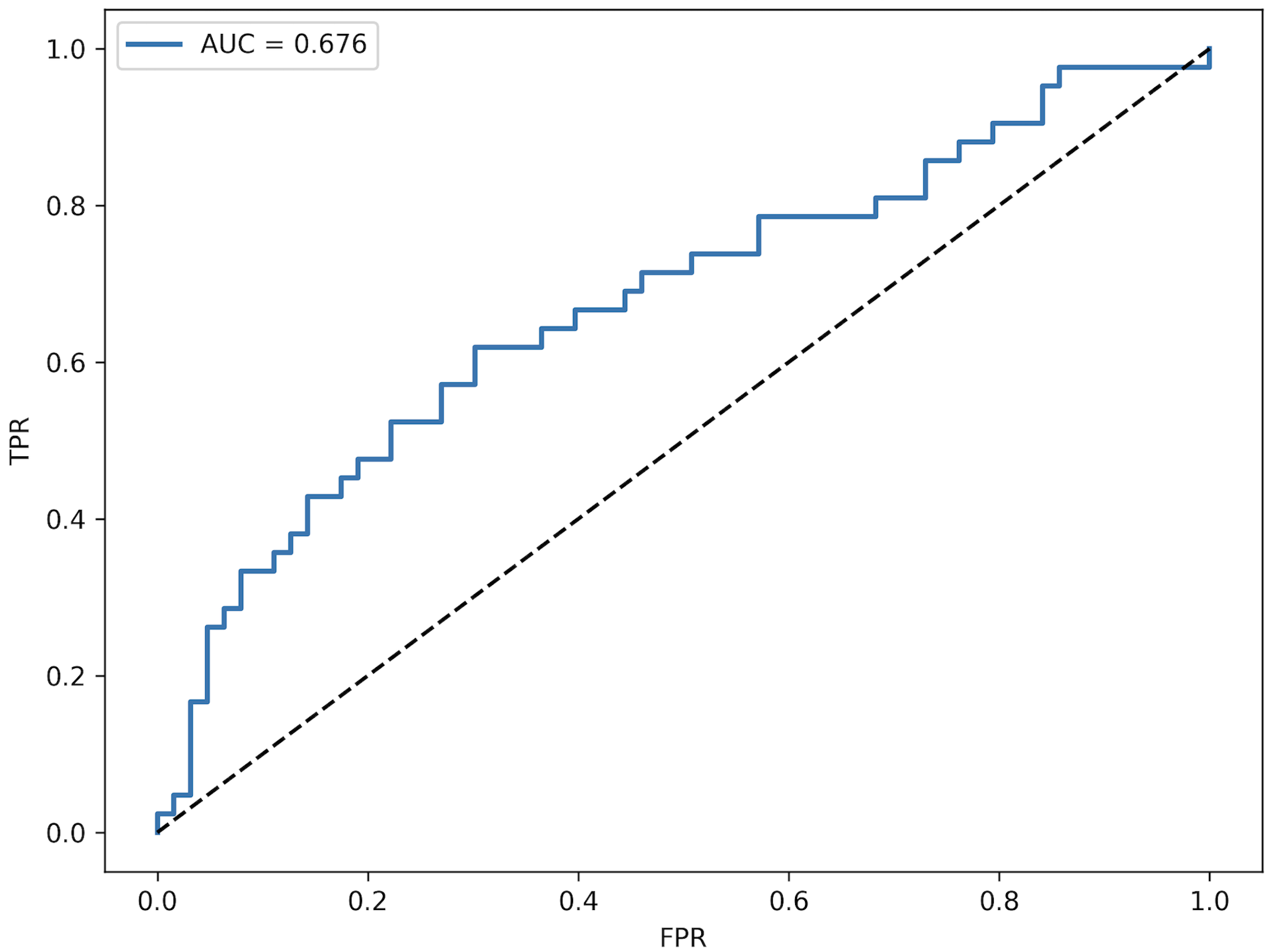}
    \caption{ROC Curve for Swin-XGBoost Ensemble.}
    \label{fig:roc}
\end{figure}

Focusing on the scores of the best performing model in table \ref{tab:scores}, Swin-XGBoost ensemble, scores show stronger performance in predicting the not effective class relative to the effective. For the not effective class the model achieves a recall of 0.85, precision of 0.68 and an F1 score of 0.75. These are strong scores and show the models ability to predict the not effective class accurately. For the effective class we see decreases in scores to 0.64, 0.44 and 0.51 for precision, recall and F1 score. These drops in scores may partly be due to the slight imbalance in the dataset pushing the model to favor the not effective class. This is further supported by the confusion matrix in table \ref{tab:confusionMatrix}, where the pipeline has a much easier time correctly classifying cases as not effective since the model predicts 54 out of 64 correct while for the effective class it only correctly classifies 18/42. Looking at figure \ref{fig:roc} the model generates an AUC of 0.676 but with a smooth curve indicating the need to accept more FPs to get more TPs. The AUC does confirm that this model is able to distinguish the signal and holds predictive power. Generally, the aggregate of these scores are respectable given the minimum input needed by the model of 4 sequence pretreatment MRI images accompanied with a small amount of clinical data. 

% \begin{figure}[htbp]
%     \centering
%     \includegraphics[width=0.95\columnwidth]{importanceFig.png}
%     \caption{Importance (Gain) of Features used to Train XGBoost}
%     \label{fig:importance}
% \end{figure}

% \begin{table}[h!]
% \centering
% \caption{Swin Ensemble Confusion Matrix}
% \label{tab:confusionMatrix}

% % --- Table Definition ---
% % We define 6 columns (4 'c' and 2 'c')
% \begin{tabular}{|c|c|c|}
% \hline
%  & Not Effective & Effective \\
% \hline
% Not Effective & 54 & 9 \\
% \hline
% Effective & 24 & 18 \\
% \hline

% \end{tabular}
% \end{table}

\begin{table}[h!]
\centering
\caption{Swin ensemble confusion matrix}
\label{tab:confusionMatrix}
\renewcommand{\arraystretch}{1.5} % Adds padding

\begin{tabular}{cc|c|c|}
  & \multicolumn{1}{c}{} & \multicolumn{2}{c}{\textbf{Predicted Class}} \\
  & \multicolumn{1}{c}{} & \multicolumn{1}{c}{Not Effective} & \multicolumn{1}{c}{Effective} \\ \cline{3-4}
  \multirow{2}{*}{\textbf{True Class}} & Not Effective & 54 & 9 \\ \cline{3-4}
  & Effective & 24 & 18 \\ \cline{3-4}
\end{tabular}
\end{table}

As seen in table \ref{tab:SHAP}, XGBoost features were analyzed for their importance to give insights into the model's decision making using SHAP. table \ref{tab:SHAP} shows the 20 most important features used by the XGBoost classifier. The two most important features based on this analysis are the age of the patient at initial diagnosis and the tumor location both of which are clinical features. Age is by far the most significant feature with a mean SHAP score of 0.99 compared to the second most significant, tumor location, of 0.33. This validates the importance of clinical information in this classification problem and we further prove this by our XGBoost models trained separately on clinical, radiomics and combined data in table \ref{tab:compareClinical}.

Training a XGBoost model solely on radiomic features does a generally poor job of classifying with a test set accuracy of 56.19\%. Clinical data on the other hand, scores 61.90\% outperforming the combined data with a score of 60.95\%. That being said, when taking the top 10 best performing validation parameters and averaging these model results the combined model scores 63.24\% compared to the clinical model with 61.90\%. This highlights the susceptibility of these XGBoost model's to changes in parameters and that although radiomic features do aid in the classification process much of the performance comes from the clinical data. Interestingly, the clinical model does not do as well on training or validation as the other two which may be explained through better generalizing than the others through reduced overfitting \ref{tab:compareClinical}.

% \begin{table}[h!]
% \centering
% \caption{Results of Top 3 XGBoost Classifiers from the Validation Set on the Testing Set}
% \label{tab:top3scores}

% % --- Table Definition ---
% % We define 6 columns (4 'c' and 2 'c')
% \begin{tabular}{|c|c|c|c|}
% \hline
% Parameter & 1st & 2nd & 3rd \\
% \hline
% Max Depths & 3 & 4 & 5 \\
% Min Child Weights & 2 & 4 & 4 \\ 
% Subsamples & 0.76 & 0.80 & 0.82 \\ 
% Colsample by Trees & 0.74 & 0.64 & 0.71\\ 
% Learning Rates & 0.03 & 0.04 & 0.04\\ 
% Alphas & 2.1 & 2.7 & 0.1 \\ 
% Lambdas & 1.7 & 1.1 & 3.3 \\ 
% Gammas & 0.2 & 0.4 & 0.1 \\
% \textbf{XGBoost Scores} & \textbf{56.19} & \textbf{62.85} & \textbf{62.85} \\
% \textbf{Swin Ensemble Scores} & \textbf{69.52} & \textbf{65.71} & \textbf{66.66} \\
% \hline

% \end{tabular}
% \end{table}

\section{CONCLUDING REMARKS}

In this paper, we introduce a pipeline that predicts chemotherapy efficacy using simple pretreatment MRI scans accompanied by clinical information. Our pipeline combines a state-of-the-art pediatric segmentation model with radiomic feature extraction and clinical data feed through an ensemble of XGBoost and Swin UNETR encoder to accurately classify the efficacy of chemotherapy. We accomplish this through using the generated 4 label segmentation consisting of ET, NET, CC and ED and classifying based solely on this segmentation and imaging using Swin UNETR and ensembling this output with XGBoost trained on combined clinical data and radiomic features extracted using the generated segmentations and imaging.

Due to the limited availability of data and the specific requirements of our model, we were unable to perform external validation. Our approach requires pretreatment MRI images with four sequences and matched clinical information, and such comprehensive datasets are extremely rare. Even in adult brain tumor research, publicly available resources typically include imaging alone without corresponding clinical variables, further limiting external validation options. Pediatric studies face additional challenges, including small cohorts, restricted data sharing, and substantial inter-institutional variability. Because of our modest sample size, we could not conduct stratified analyses by chemotherapy regimen, tumor location, or molecular subtype. Larger multi-institutional datasets will be necessary to validate these findings, enable subgroup-specific modeling, and support future prospective clinical evaluation.

Our best-performing Swin UNETR encoder combined with the XGBoost ensemble achieved a 69\% accuracy on held-out test sets using five-fold cross-validation. Performance was higher for the not effective (majority) class than for the effective class, reflecting class imbalance. Future work will focus on expanding the dataset and exploring oversampling strategies to improve minority-class performance.

As mentioned briefly in the results the XGBoost classifiers are very sensitive to changes in parameters which can alter their training and performance. In our testing when averaging over the top 1\% of best performing parameters, scores increased from the model we had chosen by approximately 1.5\%. This is not inconsequential and goes to show the the variability in these models.

To the best of our knowledge, this is the first study to predict chemotherapy response in pediatric low-grade glioma using pretreatment MRI data integrated with clinical variables: including legal sex, ethnicity, race, age at event, molecular subtype, tumor location, initial surgery status, metastatic status and site, chemotherapy regimen, treatment protocol, and specific chemotherapeutic agents, combined with radiomic features. No prior work has addressed chemotherapy-response prediction at the time of direct treatment or pLGG patients who require chemotherapy because they are not amenable to complete surgical resection. Developing such early predictive capability may enable more strategic treatment planning, support individualized therapeutic decision-making, and ultimately contribute to improved survival and event-free survival outcomes for this high-risk patient population.


%%%%%%%%%%%%%%%%%%%%%%%%%%%%%%%%%%%%%%%%%%%%%%%%%%%%%%%%%%%%%%%%%%%%%%%%%%%%%%%%

%%%%%%%%%%%%%%%%%%%%%%%%%%%%%%%%%%%%%%%%%%%%%%%%%%%%%%%%%%%%%%%%%%%%%%%%%%%%%%%%

%%%%%%%%%%%%%%%%%%%%%%%%%%%%%%%%%%%%%%%%%%%%%%%%%%%%%%%%%%%%%%%%%%%%%%%%%%%%%%%%

\section*{ACKNOWLEDGMENT}

This work is supported by Malnati Brain Tumor Initiative, Northwestern University. We would like to express our sincere gratitude to the CBTN for their exceptional data repository and invaluable support for Pediatric Brain Tumor Research. All other authors have no COI (conflict of interest).

%%%%%%%%%%%%%%%%%%%%%%%%%%%%%%%%%%%%%%%%%%%%%%%%%%%%%%%%%%%%%%%%%%%%%%%%%%%%%%%%

\bibliographystyle{IEEEbib.bst}
\bibliography{refs}

\end{document}